# Objects Moving along Closed Timelike Curves belong to Proper Classes

Zhongzhu Liu

Department of Physics, Huazhong University of Science and Technology, Wuhan 430074, People's Republic of China

**Abstract** According to the set theory, we prove that objects moving along closed timelike curves (CTCs) should belong to proper classes, but never to any set. Particles in a set have to change own some properties when they come into a CVC in order to become the objects in a proper class.

Closed timelike curves (CTCs) appear in various theories concerning the gravitation [1-8]. They destroy the causation law, so brought an interpretive difficulty into the physics [1-8]. Carter ascribed the emersion of CTCs to "whole space is a single nontrivially vicious set in which any point can be connected to other point by both a future and past directed timelike curve" [5]. People leave no stone unturned to keep the causation law in theories with CTCs violating the law of causation. But, these efforts has no brought remarkable change to this dilemma up to the present [9-12]. We should realize that all solutions to gravitation field are geometrical results. Geometry never has the causation law. This law relates only to matter, or more precisely, to the existence and movement of objects. All objects known are regarded to obey the causation law. Consequently, we understand the Carter's "vicious" as the inconsistence between properties of objects and CTCs. In this case, people have to consider the alternative possibility that we do not go to study how the causation law persecutes the movement of objects along CTCs, but go to study what properties an object will possess when it moves along CTCs. This is an idea against our custom. So, we must study such objects on the most base. The set theory and the logic are the base of all mathematics and physics [13]. Thus, we shall make the concrete analysis about objects moving along CTCs according to the set theory. And we shall give a indirect and a direct proof, respectively, that such objects will not belong to the sets familiar to us, but belong to proper classes [14-16]. Sets and proper classes are different logical configurations. Objects belonging to them have different properties. In usual movement, every particle is regarded as a member of some set. But, it should belong to a proper classe when moves along a CTC. This means that, particles have to change own some properties in order to become objects in a proper class when they are falling to a CVC.

The set theory defines collections of objects as classes [14-16]. It claims that there are two basic classes: one is the sets prescribed by the axioms of the set theory, and the other are the proper classes disobey some axioms of the set theory [14-16]. In the set theory, the Russell's paradox is expressed by the following perspective. Let X be a class, it has the property $X \in X$ [14-16]. A simple deduction of the set theory results in that a set class $X$ possesses the property $X \notin X$ as well [14-16]. Such class possesses equally the property $X \in X$ and $X \notin X$. It involves an intrinsic inconsistency. So, it does not be a set, but can but be a proper class. Or in other words, classes involving an intrinsic inconsistency are proper classes. The existent of CTCs lead to the paradox of "the time travel" [4,17]. Through an embodiment of the paradox, the grandmother's paradox, we prove that the paradox of "the time travel" is just the Shen You Ting's paradox [18]. The Shen You Ting's paradox is an extension of the Russell's paradox. At this rate, the intrinsic inconsistency in the paradox of "the time travel" is just the intrinsic inconsistency in the Russell's paradox. And, objects moving along a CTC are proved indirectly to belong to proper classes. Afterward, we shall show that the problem whether a particle moving along a CTC collides with bypast itself is just the barber's paradox, an extension of the Russell's paradox [14,17]. In this way, objects moving along CTCs are proved directly to belong to proper classes. Since sets and proper classes are definitely different classes, objects moving along CTCs never belong to sets.

Choose three points $x$, $y$ and $z$ on a CTC respectively. The physical time on this curve is the proper time $\tau$. A particle moves along the curve. It has the times $\tau(x)$, $\tau(y)$ and $\tau(z)$ when it passes through these three points in succession. The relation among these times is

$$\tau(x) < \tau(y) < \tau(z) < \tau(x) \qquad (1)$$

This inequality is just the paradox of "the time travel" [17]. It violates the causation law. An embodiment of the paradox of "the time travel" is the grandmother's paradox [17]: A man travels back in time and impregnates his grandmother. The result is a line of offspring and descendants, including the man's parent(s) and the man himself. Therefore, unless he makes the time-travel trip at all, he will never exist.

We can rewrite the paradox with the language of the set theory. The man's family has three hierarchies. The first one is the grandmother; the second is the man's mother and third is the man himself. We suppose all descendants of the i-th hierarchy composes the collection $X_i$ such that every of sons and daughters is a member of $X_i$, and every collections of descendants of sons and daughters $X_{i+1}$ is also a member of $X_i$. So, the man and the collections of the man's offspring belongs to $X_2$, the man's mother and the collection $X_2$ belongs to $X_1$ and the man's offspring belong to $X_3$. These collections all are classes. Therefore, suppose that apparently, we have the relations among them as follows

$$X_3 \in X_2 \in X_1. \tag{2}$$

Moreover, we have also

$$X_1 \in X_3, \tag{3}$$

because the man makes his grandmother pregnant. Combining the above two formulas, we get

$$X_3 \in X_2 \in X_1 \in X_3 \tag{4}$$

This is just the Shen You Ting's paradox in the set theory [18], an extension of the Russell's paradox. These collections do not obey the regularity axiom [14,17]. They are not sets, but are proper classes. This formula shows that the inconsistency involved in Eq.1 is just the inconsistency involved in the Russell's paradox. Accordingly, if objects possess the property represented by the Eq.1, their collections will possess equal property represented by the Russell's paradox, and these objects must belong to proper classes [19,20].

Now we carry out the direct test. A particle starts from the point $x$ on a CTC at $\tau(x)$. Going around the curve one loop, the proper time of the particle should increase $\Delta \tau$. However, its time will not change when it retours. Thus an inconsistency appears. We adopt $p(\tau)$ to denote the particle at $\tau(x)$, and $p(\tau + \Delta \tau)$ to denote it going around one loop. The above inconsistency becomes the problem whether $p(\tau + \Delta \tau)$ can collide with $p(\tau)$ [4]? In term of the acceptable definition about the particle collision, a particle can not collide with itself. On the other hand, objects at the same spacetime point will inevitably collide. Hence an inconsistency arises. We can express the inconsistency more explicitly by emphasizing that any particle can not collide with itself. Thus, we define the collision among particles as that a particle can and only can collide with all particles which never collide with themselves. The inconsistency emerges when a particle and bypast itself appear at the same spacetime point. Here, if $p(\tau + \Delta \tau)$ can collide with $p(\tau)$, then $p(\tau + \Delta \tau)$ can not collide with $p(\tau)$ according to the definition! Or, if $p(\tau + \Delta \tau)$ can not collide with $p(\tau)$, and then $p(\tau + \Delta \tau)$ can collide with $p(\tau)$ correspondingly!

The Russell's paradox has an embodiment, the barber's paradox: "there is a town with a barber who shaves all the people (and only the people) who don't shave themselves" [6,7]. Comparing the definition about the collision among particles to the barber's paradox, we can conclude that they share the identical logistic configuration. So, if a object is able to move along CTC, it certainly belongs to the collections possessing the property expressed by the Russell's paradox. such a collection is a proper classe.

In the set theory, the Russell's paradox is appearing when prescribe a class by the Comprehension principle [9]. Noteworthily, the preper classes containing objects rounding along CTCs are prescribed no by the Comprehension principle, but by the speciality of the spacetime. A paradox introduced by adopting the Comprehension principle can be canceled by a accessional rule [9], and the preper classes disappear accordingly. However, the paradox prescribed by physical processes are irrepealable to disappear unless the processes change. So, such preper classes will not disappear for the motion along CTCs. And, people can not change a such preper class into a set.

Specific particles obey specific physical laws. We've already known all physical laws of particles studied by us. Mathematics describing these laws, whether the analyse or the algebra (except the homotopical theory [21]), all are constructed on the basis of the set theory [22,23]. So, all the theory describing known particle collections is the set theory. Whereas, proper classes can not be described by the set theory because they have their own mathematical base [14-21].

An object can and only can belongs to either sets or proper classes. Objects in a set should possesses some physical properties different from that obeyed by objects in a proper class. A particle moving along a causal trajectory belongs to a set. If this particle is propelled by some external effect to fall on a CTC, its some properties must change, because the particle set is becoming a proper class along with the fall.

To sum up, objects moving along CTCs belong to proper classes. Moreover, objects that are capable of the time travel merely belong to a proper class, but never to any specific set. Therefore, particles in a set have to change own some properties when they come into a CVC in order to become objects in a proper class.